\documentclass[epj]{svjour}
\usepackage{graphics}
\begin{document}
\title{Glueballs and vector mesons at NICA}
\author{Denis Parganlija\inst{} 
}                     
%
%
\institute{Technische Universit\"{a}t Wien, Institut f\"{u}r Theoretische Physik, Wiedner Hauptstr.\ 8-10, 1040 Vienna, Austria 
}
\date{16.10.2015 / Revised version: 07.01.2016}
%
\abstract{
Two interconnected fields of interest are suggested for NICA.
Firstly, existence of glueballs is predicted by the theory of strong interaction
but -- even after decades of research -- glueball identification in
the physical spectrum is still unclear. NICA can help to ascertain experimental
glueball candidates via $J/\Psi$ decays whose yield is expected to be large.
Importance of glueballs
is not limited to vacuum: since they couple to other meson states, glueballs can also be expected to influence signatures of
chiral-symmetry restoration in the high-energy phase of strong dynamics. 
Mass shifting or in-medium broadening of vector and axial-vector mesons may occur there but the extent of such phenomena
is still uncertain. Additionally, glueball properties could also be modified in medium. Exploration
of these issues is the second suggested field of interest that can be
pursued at NICA.
} 
\maketitle

\section{Introduction}
\label{Introduction}

Quantum Chromodynamics (QCD), the established theory of the strong
interaction, is per construction of non-Abelian nature. As a consequence,
gauge bosons of QCD -- the gluons -- are self-interacting. Since the strong
coupling is large at sufficiently small energies \cite{af1,af2}, the expectation
is that the non-perturbative region of strong dynamics enables gluons to
build more complex objects denoted as glueballs \cite{glueballreferences1,glueballreferences2,glueballreferences3,glueballreferences4,glueballreferences5,glueballreferences6,glueballreferences7,glueballreferences8}.
Theoretical studies have shown lively interest in glueballs using various
methods to approach the non-perturbative regime of QCD:

\begin{itemize}

\item Ab-initio numerical calculations in lattice QCD have resulted in predictions of glueball spectra in 
quenched as well as unquenched approximations \cite{Morningstar,Lee,Bali:2000vr,Hart,Loan,Gregory2006,Chen,Hart:2006ps,Richards,Gregory,Yang}.

\item The AdS/CFT correspondence has yielded results both on glueball spectra \cite{BMT,Capossoli:2013kb,Capossoli:2015ywa,Capossoli:2016kcr,Capossoli:2016ydo}
as well as decays of glueballs \cite{Hashimoto:2007ze,Brunner:2015oqa,Brunner:2015yha,Brunner:2015oga}; an exemplary
approach in this direction is briefly outlined in Sect.~\ref{Sec.2}.

\item Effective approaches to QCD have upon implementation of relevant symmetries of strong dynamics
considered not only glueball decays but also various mixing mechanisms between glueball and non-glueball states
obtaining a satisfactory overall agreement with experimental data
\cite{glueballmodels,glueballmodels1,glueballmodels2,glueballmodels3,glueballmodels4,glueballmodels5,Janowski:2011gt,Janowski:2014ppa}; see also Refs.\ \cite{Narison:1988ts,Narison:1996fm,Sanchis-Alepuz:2015hma}.

\end{itemize}
$\,$ \\
There are several reasons for interest in glueballs:

\begin{itemize}

\item Glueballs are unique since their mass is, at the leading order, generated solely via self-interaction of gluons
(pure gluodynamics). Although at the level of full QCD current quark masses can contribute, their effects are currently unclear.
This is particularly the case in lattice QCD where the inclusion of dynamical fermions leads to the emergence 
of states additional to those present in pure gluodynamics with consequences that, \textit{e.g.},
(\textit{i}) states experience overlaps and (\textit{ii}) the scalar glueball is no longer
the lowest state of the spectrum. Then the identification
of states is more complicated -- and conclusions from lattice QCD in the scalar channel
somewhat conflicting: Refs.\ \cite{Gregory2006,Richards,Gregory} do not observe large unquenching effects in simulations
relying on staggered fermions while a different opinion (in line with
the expectation of Ref.\ \cite{Loan}) is advocated by
simulations with clover fermions \cite{Hart:2006ps}.\\
Leading-order mass generation of glueballs is in contrast to other strongly interacting particles (\textit{i.e.}, hadrons) whose masses are
predominantly generated by quark dynamics and thus susceptible to, albeit very small, contribution of the 
Brout-Englert-Higgs mechanism (see examples for pions \cite{GellMann:1968rz}; kaons \cite{Shifman:1978bw}; $\omega$-$\rho$ splitting
\cite{Shifman:1978bw,Shifman:1978by}; nucleons \cite{Procura:2006bj}).\\
Therefore glueballs represent a very important tool to explore strong dynamics.

\item The spin of glueballs is integer since gluons are vector particles. Consequently the spectrum of mesons
(\textit{i.e.}, hadrons of integer spin) would be incomplete if glueballs were omitted from experimental searches.

\end{itemize}
In Sect.~\ref{Sec.3}, some of the issues on the experimental side of the glueball search are exemplified,
together with suggestions for NICA in this regard.
\\
\\
Already in vacuum, glueballs couple to non-glueball states 
(possessing $\bar{q}q$, $\bar{q}\bar{q}qq$ and other valence degrees of freedom). 
The expectation is that a coupling of modified strength will remain at non-zero temperatures and densities. In that case,
glueballs will influence $\bar{q}q$ states and the underlying phenomena of their in-medium behaviour, such as the
chiral-symmetry restoration. It is, however, unclear what this behaviour
exactly entails since vector and axial-vector mesons may shift in mass
or become broader in medium but clear experimental evidence for this is still outstanding.
These issues together with further suggestions for NICA are
discussed in Sect.~\ref{Sec.4}. Conclusions are presented in Sect.~\ref{Conclusions}.

\section{Hallmarks of a glueball: an example}
\label{Sec.2}

A glueball state can be distinguished from other hadrons by for example
(\textit{i}) strong suppression in two-photon decay channels \cite{Wiedner:2013bta} and prominent
presence in radiative decays \cite{Yang}; (\textit{ii}) decay patterns. Various approaches to glueball
dynamics (mentioned in the previous section) have been applied in studies of glueball decays;
in the following, a recent approach based on the AdS/CFT correspondence is briefly discussed
and its results for glueball identification are presented.
\\
\\
The approach is based on the conjectured duality between weakly coupled string theory (\textit{i.e.}, supergravity)
in an anti-de Sitter (AdS) space and a strongly coupled conformal field theory (CFT) in one dimension less
\cite{Maldacena}. The field theory possesses symmetries absent from QCD (supersymmetry in addition to conformality); these are removed by suitable compactifications in the full supergravity
space \cite{Witten} and the emerging $U(N_c)$ gauge theory (with $N_c \rightarrow \infty$) may be used to explore the
Yang-Mills sector of QCD. Then (holographic) glueballs are obtained as graviton polarisations in the supergravity
background. It was demonstrated in Ref.\ \cite{BMT} that such an approach leads to a glueball spectrum
that is remarkably similar to the one obtained in lattice-QCD simulations.
\\\\
Studying glueball decays into $\bar{q}q$ states requires introduction of quark degrees of freedom.
A method to include chiral quarks -- the so-called Witten-Sakai-Sugimoto (WSS) Model -- was proposed in
Refs.\ \cite{SaSu1,SaSu2} by introducing $N_f$ (number of flavours) 
probe D8- and anti-D8-branes in the supergravity space that extend along all dimensions in the space except for a 
(Kaluza-Klein) circle. 
D-branes introduce a $U(N_f) \times U(N_f)$ symmetry in the theory; since D8- and anti-D8-branes merge at a 
certain point in the bulk space, 
the original $U(N_f) \times U(N_f)$ symmetry is reduced to its diagonal subgroup. This is interpreted 
as a geometric realisation of chiral-symmetry breaking.
\\
\\
It was demonstrated already in Refs.\ \cite{SaSu1,SaSu2} that
the WSS Model can describe phenomenology
of $\bar{q}q$ states at least in a semiquantitatively correct way. Decays of dilaton glueballs
were in turn explored in Refs.\ \cite{Brunner:2015oqa,Brunner:2015yha} 
where predictions for decays of the scalar and tensor glueballs in the $2\pi$, $4\pi$, $6\pi$, $2K$
and $2\eta$ channels have been made, as presented in Tables \ref{Table1} and \ref{Table2}.
\\
\\
However, irrespective of the lively theoretical interest
in glueballs, the identification of these states in the physical spectrum is still outstanding.

\begin{table}[b]
\caption{Comparison of holographic scalar-glueball decays
(the outer right column) obtained in Ref.\ \cite{Brunner:2015oqa} with experimental data for the two prime
candidates for the scalar glueball, the resonances $f_0(1500)$ and $f_0(1710)$. The 't Hooft coupling 
(that is the only free quantity in the WSS Model, except for the Kaluza-Klein mass
which sets the Model scale) was determined in two ways,
by implementing the experimental value of the pion decay constant
or the lattice-QCD value of the string tension. 
This allows for theoretical uncertainties to be estimated and hence holographic results are 
presented in intervals. 
All experimental data
are from PDG \cite{PDG} except for those marked by a star that are from
Ref.\ \cite{DA} where the $f_0(1710)$ decay channels were calculated assuming a negligible coupling of that resonance to $4\pi$. All masses are in MeV. 
The $f_0(1710)$ resonance is preferred to have a significant overlap
with the scalar glueball but a conclusive statement in this regard is 
hampered by experimental uncertainties discussed in Sect.~\ref{Sec.3}.}
\label{Table1} 
\begin{tabular}{llll}
\hline\noalign{\smallskip}
Decay & $M_{\mbox{exp.}}$ & $\Gamma/M$ (exp.) & $\Gamma/M$ (holography)  \\
\noalign{\smallskip}\hline\noalign{\smallskip}
$f_0(1500)$ (total) & 1505 & 0.072(5) & 0.027...0.037  \\
$f_0(1500) \rightarrow 4\pi$ & 1505 & 0.036(3) & 0.003...0.005  \\
$f_0(1500) \rightarrow 2\pi$ & 1505 & 0.025(2) & 0.009...0.012 \\
$f_0(1500) \rightarrow 2K$ & 1505 & 0.006(1) & 0.012...0.016 \\
$f_0(1500) \rightarrow 2\eta$ & 1505 & 0.004(1) & 0.003...0.004 \\
\noalign{\smallskip}\hline
$f_0(1710)$ (total) & 1723 & 0.078(4) & 0.059...0.076 \\
$f_0(1710) \rightarrow 2K$ & 1723 & 0.047(17)* & 0.012...0.016 \\
$f_0(1710) \rightarrow 2\eta$ & 1723 & 0.022(11)* & 0.003...0.004 \\
$f_0(1710) \rightarrow 2\pi$ & 1723 & 0.009(2)* & 0.009...0.012 \\
$f_0(1710) \rightarrow 4\pi$ & 1723 & ? & 0.024...0.030 \\
$f_0(1710) \rightarrow 2\omega \rightarrow 6\pi$ & 1723 & seen & 0.011...0.014 \\
\noalign{\smallskip}\hline
\end{tabular}
\end{table}

\begin{table}[h]
\caption{Decays of the holographic tensor glueball predicted by the
WSS Model
for two different masses, $M_T = 2000$ MeV and $M_T = 2400$ MeV \cite{Brunner:2015oqa}.
The former mass is chosen to approximately correspond to that of
the $f_2(1950)$ resonance, a possible candidate for the tensor glueball
due to its mostly flavour-blind decay modes; for this state, $\Gamma/M=0.24(1)$ where $\Gamma$ is the total decay width \cite{PDG}. The value 
$M_T = 2400$ MeV is chosen exemplary as an element of the intervall for the tensor-glueball mass
predicted by lattice QCD \cite{Morningstar,Hart,Loan,Chen,Gregory}.
Just as for results presented in Table \ref{Table1}, the 't Hooft coupling was determined in two ways:
by implementing the experimental value of the pion decay constant
or the lattice-QCD value of the string tension. Holographic results are 
thus presented in intervals
in order to estimate theoretical uncertainties. }
\label{Table2}
\begin{tabular}{lll}
\hline\noalign{\smallskip}
Decay & $M_T$ (MeV) & $\Gamma/M_T $ (holography)   \\
\noalign{\smallskip}\hline\noalign{\smallskip}
$T \rightarrow 2\rho \rightarrow 4\pi$ & 2000 & 0.135...0.178  \\
$T \rightarrow K^{*} K^{*} \rightarrow 2(K \pi) $ & 2000 & 0.119...0.177  \\ 
$T \rightarrow 2\omega \rightarrow 6\pi$ & 2000 & 0.045...0.059  \\
$T \rightarrow 2\pi$ & 2000 & 0.014...0.018 \\
$T \rightarrow 2K$ & 2000 & 0.010...0.013  \\
$T \rightarrow 2\eta$ & 2000 & 0.0018...0.0024  \\
$T$ (total) & 2000 & $\approx$ 0.32...0.45  \\
\noalign{\smallskip}\hline
$T \rightarrow K^{*} K^{*} \rightarrow 2(K \pi) $ & 2400 & 0.173...0.250  \\ 
$T \rightarrow 2\rho \rightarrow 4\pi$ & 2400 & 0.159...0.211  \\
$T \rightarrow 2\omega \rightarrow 6\pi$ & 2400 & 0.053...0.070 \\
$T \rightarrow 2\phi$ & 2400 & 0.032...0.051  \\ 
$T \rightarrow 2\pi$ & 2400 & 0.014...0.019 \\
$T \rightarrow 2K$ & 2400 & 0.012...0.016  \\
$T \rightarrow 2\eta$ & 2400 & 0.0025...0.0034  \\
$T$ (total) & 2400 & $\approx$ 0.45...0.62  \\
\noalign{\smallskip}\hline
\end{tabular}
\end{table}

\section{Experimental ambiguities relevant for glueballs: an example,
and a suggestion for NICA}
\label{Sec.3}

Reasons for problems in experimental identification of glueballs
are at least twofold:

\begin{itemize}

\item Glueballs are expected to emerge starting at energies between
approximately 1.5 GeV and 1.8 GeV where the ground state, a scalar \cite%
{West}, is predicted in numerical simulations of the spectrum \cite{Morningstar}. 
Historically there has been a scarcity of precise
experimental data exactly in the energy region where glueballs are
expected to emerge \cite{PDG}. Although there has been a notable change in
data availability \cite{BES,Belle,LHCb}, the amount of progress is still not
sufficient for an unambiguous identification of these states.

\item Glueball with a given set 
of quantum numbers will inevitably
mix/interfere with non-glueball states (possessing $\bar{q}q$, $\bar{q}\bar{%
q}qq$ and other valence degrees of freedom) that have the same quantum
numbers. The effects of such interference in experimental data render the
identification of resonances in general, and thus glueballs in particular,
highly non-trivial \cite{Bugg:2004xu}.

\end{itemize}
$\,$\\
Existing issues in experimental glueball searches can be illustrated by the following example
relevant for the scalar glueball. This state possesses quantum numbers
$IJ^{PC}=00^{++}$ where $I$, $J$, $P$ and $C$ respectively denote the isospin, total
spin, parity and charge conjugation.
Particle Data Group (PDG) cites the existence of five $IJ^{PC}=00^{++}$ resonances in the energy
region up to $\sim$ 1.8 GeV:
$f_{0}(500)$, $f_{0}(980)$, $%
f_{0}(1370)$, $f_{0}(1500)$ and $f_{0}(1710)$. They are known as scalar
isoscalar resonances \cite{PDG}; for a brief review, see Refs.\ \cite{Parganlija:2012nc,Parganlija:2013xsa}. 
Claims have been made \cite{1790-1,1790-2,BESII2004-1,BESII2004-2}
that a sixth such state exists, namely $f_{0}(1790)$ -- a state very close
to $f_{0}(1710)$ but with a different decay behaviour: $f_{0}(1790)$ decays
predominantly into pions whereas $f_{0}(1710)$ decays predominantly into
kaons.
\\
\\
There are four basic production mechanisms for $f_0(1710)$ and $f_0(1790)$
via $J/\psi$ decays:

\begin{itemize}
\item (\textit{i}) $J/\psi\rightarrow\phi K^+ K^-$,

\item (\textit{ii}) $J/\psi\rightarrow\phi \pi^{+}\pi^{-}$,

\item (\textit{iii}) $J/\psi\rightarrow\omega K^{+}K^{-}$,

\item (\textit{iv}) $J/\psi\rightarrow\omega \pi^{+}\pi^{-}$.
\end{itemize}
Reactions (\textit{i}) and (\textit{iii}) allow for reconstruction of $%
f_{0}(1710)$ -- see Ref.\ \cite{1710} -- whereas $f_{0}(1790)$ is reconstructed from reactions (\textit{ii}) and
(\textit{iv}). Importantly,
assuming $f_{0}(1710)$ and $f_{0}(1790)$ to be the same resonance leads to a
contradiction: such a resonance would have to possess a pion-to-kaon-decay
ratio of $1.82 \pm 0.33$ according to reactions (\textit{i}) and (\textit{ii}%
) and a pion-to-kaon-decay ratio $< 0.11$ according to reactions (\textit{%
iii}) and (\textit{iv}) \cite{BESII2004-1,BESII2004-2}.
Decay ratios must be
independent of the production mechanism for a single resonance. 
The assumption that  $f_{0}(1710)$ and $f_{0}(1790)$ represent a single resonance clearly leads to a contradiction in the value of the
mentioned decay ratio; consequently, the employed data -- obtained by the BES Collaboration -- prefer $f_{0}(1790)$
as a resonance distinct from $f_0(1710)$. Nonetheless, additional inspection of this claim is by all means needed in further
experiments.
\\
\\
If the existence of the $f_{0}(1790)$ resonance is confirmed, it will most certainly have implications
for glueball search since its mass is within the interval in which the scalar glueball is expected
to appear according to lattice QCD.
\\
\\
NICA \cite{NICA} program focused on the Spin Physics Detector (SPD) \cite{SPD,SPD1} appears to be relevant for the issue of $f_{0}(1790)$ but could
also discover further resonances.
If the Monte Carlo simulations of $J/\Psi$ production rates at SPD prove correct, then the yearly yield of these resonances
should amount to $\sim 10^7$ events \cite{SPD,SPD1}. It
would thus be of the same magnitude as that of BES II \cite{BESII2004-1,BESII2004-2}, where the best available
evidence for the existence of $f_0(1790)$ has been presented -- 
and then a careful reconstruction of resonances
in $2\pi$ final states emerging from $J/\Psi$ decays could clarify whether $f_0(1790)$ exists. New resonances even higher in energy may also be discovered or those for which there is already indication
-- $f_0(2020)$, $f_0(2100)$, $f_0(2200)$, $f_0(2330)$ \cite{PDG} --
could be confirmed.
\\
\\
Comparison of SPD with other running or planned programs is in order. Given the above data on $f_0(1790)$, two sorts of production mechanisms
are particularly relevant: (\textit{i}) $e^+ e^-$ (as at BES) and (\textit{ii}) $pp$ (since planned at SPD).
\\
\\
Firstly, $e^+ e^-$ collisions at the VEPP-4M Collider have produced $\sim$ 7 million $J/\Psi$ events, as reported by the KEDR Collaboration
\cite{BALDIN:2014wia}. This could in principle enable the reconstruction of $f_0(1790)$ but an even larger $J/\Psi$ yield is expected
at SPD.\\
Additionally, CMD-3 and SND Collaborations at VEPP-2000 can
use $e^+ e^-$ collisions for scans of the energy region from hadron-production
threshold up to 2 GeV but their focus is currently on vector mesons
only \cite{Schwarz:2014nfa}, and SPD could fill this gap.\\
Note further that, although the primary focus of Belle-II \cite{Abe:2010gxa}
is on precision measurements beyond the Standard Model, discoveries
in non-perturbative QCD can be expected also from that source given
the large expected luminosity (larger than at SPD). Belle-II
will rely on reconstruction of resonances from $\Upsilon(4S)$ rather than
$J/\Psi$ decays; particles below 2 GeV may be nonetheless reconstructable
but, given the large difference in mass and the well-known issues of
overlapping scalar states, great care will have to be given to proper data
analysis. Lower statistics should be sufficient for SPD to reach the same
goal since the $J/\Psi$ production is expected to be abundant.
\\
\\
Proton-proton collisions are nowadays most prominent at the LHC.
It is clear that the LHCb \cite{LHCb}, TOTEM \cite{Anelli:2008zza} and
ALICE \cite{Lamsa:2010fe} Collaborations can draw on huge cross-sections
obtained at very large energies. Nonetheless, a comparative advantage
of the SPD program is the use of polarised proton and deuteron beams
that were of enormous importance for meson discoveries in the past
\cite{Bugg:2004xu}.
\\
\\
Historically, proton-proton collisions have always represented a 
method of meson research with a large discovery potential even
at moderate beam energies \cite{DA} that has proven complementary to 
antiproton-proton \cite{PANDA} and pion-nucleon \cite{COMPASS} collisions or to 
photoproduction \cite{GlueX,CLAS12}.
\\
\\
My suggestion is thus that NICA Collaboration measure at SPD the number of events as a function of centre-of-mass energy
for $2\pi$, $2K$ and $4\pi$ final states at energies above $\sim $ 1.5 GeV and carefully analyse the data for new resonances. Further final states can be
analysed as the data become available. Glueball production may be copious
in any of these channels, with results presented in Tables \ref{Table1} and \ref{Table2} suggesting a very prominent
$4\pi$ coupling both for scalar and tensor glueballs. The potential for the discovery of new resonances thus appears to be large,
with consequences even for non-glueball states. 

\section{Vector mesons and NICA}
\label{Sec.4}

There are four established resonances in the $J^{PC} = 1^{--}$ (\textit{i.e.}, vector) meson
channel in the energy region up to approximately 1 GeV: $\rho(770)$, $\omega(782)$,
$K^{*}(892)$ and $\phi(1020)$ \cite{PDG}. In the $J^{PC} = 1^{++}$ (\textit{i.e.}, axial-vector) channel,
the established resonances up to 1.5 GeV are $a_1(1260)$, $f_1(1285)$, $K_1(1270)$, $K_1(1400)$ and $f_1(1420)$ with the $K_1$ states possibly having noticeable admixture
from the $J^{PC} = 1^{+-}$ (pseudovector) channel \cite{Divotgey:2013jba}.
Phenomenology of all of these resonances has been extensively studied in vacuum -- see Refs.\ \cite{Parganlija:2010fz,Parganlija:2012fy}
and refs.\ therein; chiral partners among these resonances can offer insight into important
phenomena of high-energy QCD \cite{RS}.
\\
\\
Current experimental ambiguities regarding mesons in medium can be illustrated by the following example:
from the side of theory, general expectation is that 
the (axial-)vector masses will follow one of the following two scenarios:

\begin{itemize}
\item The mass decreases to zero as the chiral condensate vanishes (\textit{i.e.},
the chiral symmetry of QCD becomes restored) -- the "Brown-Rho scenario" \cite{BR}.

\item The mass remains essentially constant or decreases marginally as the
chiral condensate vanishes -- the so-called "constant-rho scenario" \cite{RS}.
\end{itemize}
The constant-rho scenario is based on an obervation, \textit{e.g.}, from Linear Sigma Model with vector and axial-vector mesons
that the $\rho$ meson -- although consistent with a $\bar{q}q$ state
\cite{Pelaez:2006nj} -- actually has two contributions to its mass, one from the chiral and another from the
gluon condensate; an overall meson study \cite{Janowski:2011gt,Parganlija:2010fz,Parganlija:2012fy} then suggests that $m_\rho$ is dominated by the gluon condensate
rather than by the chiral one.
\footnote{It has to be noted here that the mentioned expectation is based on experimental data in vacuum that suffer from uncertainties discussed
in Sect.~\ref{Sec.3}. For this reason, improved measurements in vacuum physics would enable
more precise theoretical predictions of in-medium meson properties.}
\\
\\
Two questions are crucial: (\textit{i}) the behaviour of the gluon condensate in medium;
(\textit{ii}) the behaviour of vector mesons in medium.
\\
\\
General conclusion from a range of approaches is that the gluon condensate is virtually
unchanged below a critical temperature $T_c$
whose value in lattice QCD is strongly dependent on whether pure gluodynamics is considered
or, in addition, effects of massive quarks.
For pure Yang-Mills QCD, there is a sharp drop of the gluon condensate
at $T_c \simeq 260$ MeV; however, if light quarks are present then the
condensate exhibits a more gradual decrease between temperatures
of $\simeq 130$ MeV and $\simeq 190$ MeV
\cite{Boyd:1996bx,Boyd:1996ex,Miller:2000ki,Miller:2006hr}.
Effective models of QCD have found the gluon condensate to remain
stable up to
$T \simeq 200$ MeV \cite{Sollfrank:1994du,Agasian:1997zr,Sasaki:2011sd};
see also Refs.\ \cite{Agasian:1993fn,Volkov:1994eu}. A similar result has been obtained
from finite-$T$ renormalisation group equations \cite{Schaefer:2001cn}.
\\
\\
Measurements of in-medium vector mesons have so far obtained conflicting results on the
issue of mass shift but also on the related question of whether these resonances
experience an in-medium broadening \cite{ELSA,KEK1,KEK2,CLAS,PHENIX,ANKE,MAMI-C}.
\\
\\
High-energy limit of QCD is dominated by a gluon-rich
environment \cite{McLerran:2009ry,McLerran:2014hza,Akiba:2015jwa}.
It is therefore quite possible that glueballs influence phenomena
emerging in this phase of QCD, particularly scalar and tensor ones \cite{Buisseret:2009eb}.
Lattice simulations in pure Yang-Mills
QCD have found the scalar state to exhibit
a mass decrease starting at $T \geq 200$ MeV with a mass drop of
approximately 300 MeV and a thermal decay width
of $\sim 300$ MeV at $T = T_c$ \cite{Ishii:2002ww}. The tensor mass
is claimed to decrease only slightly below $\sim 200$ MeV but, once
$T_c$ is reached, the mass drops by approximately 500 MeV and a thermal width
of $\simeq 400$ MeV is obtained. Similar results were obtained in lattice
simulations presented in Ref.\ \cite{Meng:2009hh}.\\
Additionally, T-matrix formalism of Ref.\ \cite{Lacroix:2012pt} finds the scalar glueball
to start dissolving at $T\sim (1.3-1.5) T_c$ while the dissolution onset for the tensor
is at $T\sim 1.15 T_c$. As the temperature increases, 
the scalar glueball
becomes massless at $T \sim 900$ MeV according to Ref.\  \cite{Kochelev:2015rqa}; see also Refs.\ \cite{Vento:2006wh,Kochelev:2006sx}.
\\
\\
Hence there are many theoretical predictions, and new experimental measurements are needed.
\\
\\
Two of the planned experiments at NICA appear to be relevant here:
(\textit{i}) MultiPurpose Detector (MPD) program
intended to study hot and dense baryonic matter in heavy-ion collisions at a centre-of-mass energy up to
11 GeV \cite{MPD,MPD1} and (\textit{ii}) Baryonic Matter at Nuclotron (BM@N),
focused on production of strange matter in heavy-ion collisions at beam energies between
2 AGeV and 6 AGeV \cite{BMN,BMN1}.
My suggestion is that
NICA Collaboration perform a careful study of in-medium spectral functions
for vector and axial-vector mesons listed at the beginning of this section
-- in this way information can be obtained on the mass shifts, decay properties and other phenomena
that can improve theoretical studies of chiral-symmetry restoration.
\\
\\
A range of measurements has already
been performed at RHIC \cite{RHIC} and LHC \cite{ALICE} exploring high temperatures and low baryon densities
and at HADES \cite{HADES} exploring lower temperatures and moderate densities. The main
interest of NICA/MPD and BM@N is in the region of QCD phase diagram 
intermediate to the mentioned two, building on the results obtained at SPS
\cite{SPS}.
Hence future measurements at NICA appear to open a unique possibility to study
in particular (axial-)vector mesons at high densities and moderate
temperatures. The Collaboration also estimates that collider experiments
at MPD will have a nearly constant acceptance and
occupancy, unlike the future FAIR/CBM experiment \cite{CBM}
that will rely on a fixed target.
Exploration of (axial-)vectors under these conditions is
obviously highly desirable.
\\
\\
As an example, the degeneration of the chiral partners $\rho$ and $a_1$
can be used as an order parameter
for the chiral transition (see Refs.\ \cite{RS,Hohler:2013eba,Hohler:2015iba,Kovacs:2016juc} and refs.\ therein).
Then there are three possible
scenarios for the mass shifts of $\rho$ and $a_1$ in medium: (\textit{i}) both masses decrease and become degenerate;
(\textit{ii}) both masses increase and become degenerate; (\textit{iii}) $%
m_\rho$ increases and $m_{a_1}$ decreases leading to the degeneration of the
two masses. Currently it is unclear which of these options is realised in strong dynamics and MPD/BM@N data could provide 
valuable information in this direction.\\
Note, however, that the physical $\rho$ meson
has also been suggested to represent a superposition of states whose chiral partners are, respectively, an axial-vector and and a pseudovector 
\cite{Glozman:2003bt,Glozman:2007ek}. Patterns of chiral-symmetry restoration may be more complicated in this case.
Nonetheless, all these theoretical
calculations may be refined by experimental data resulting in a significantly
deeper understanding of high-energy QCD.

\section{Conclusions}
\label{Conclusions}

There are many open questions in strong dynamics at present, out of wich I have discussed two that appear
to be relevant for NICA: glueballs and (axial-)vector mesons in vacuum and in medium.\\
Glueballs, although theoretically expected to emerge as bound states of gluons in the low-energy region
of QCD, have remained elusive even after decades of research. One of the main reasons is a lack of precise experimental data.
Glueball search would be aided greatly if SPD @ NICA were to measure $2\pi$, $2K$ and $4\pi$ (and other) final states in the energy region
where glueballs are expected to start emerging, \textit{i.e.}, above $\sim 1.5$ GeV. \\
These measurements regarding vacuum strong dynamics
would have wider implications: since glueballs couple to $\bar{q}q$ states already in vacuum
they can be expected to influence $\bar{q}q$ in-medium dynamics as well. Consequently, clearer data on glueballs
in vacuum will permit a more precise prediction of dynamics at non-zero temperatures and densities -- including
chiral-symmetry restoration -- where additional ambiguities are present, particularly regarding the behaviour
of vector and axial-vector mesons such as mass shifts and in-medium broadening.
Currently the possible in-medium modifications of glueballs are also unclear.
Resolution of these questions can be aided by precise measurements at
MPD and BM@N. Thus the entire NICA project appears to have a large potential to decisively
increase our understanding of strong dynamics.
\\\\
{\bf Acknowledgments.}
I am grateful to F.~Br\"{u}nner, D.~Bugg, F.~Giacosa and A.~Rebhan for extensive discussions.
This work is supported by the Austrian Science Fund FWF, project no.\ P26366.

%
%
%

\begin{thebibliography}{}
%
%

\bibitem{af1}  D.~J.~Gross and F.~Wilczek,  
Phys.\ Rev.\ Lett.\ \textbf{30}, 1343 (1973). 
\bibitem{af2} 
H.~D.~Politzer,  
Phys.\ Rev.\ Lett.\ \textbf{30}, 1346 (1973).  

\bibitem{glueballreferences1} H.~Fritzsch and M.~Gell-Mann,  
eConf C \textbf{720906V2}, 135 (1972)  [hep-ph/0208010].
\bibitem{glueballreferences2} H.~Fritzsch and
P.~Minkowski,  
Nuovo Cim.\ A \textbf{30}, 393 (1975). 
\bibitem{glueballreferences3} R.~L.~Jaffe and K.~Johnson, 
Phys.\ Lett.\ B \textbf{60}, 201 (1976). 
\bibitem{glueballreferences4}
R.~Konoplich and M.~Shchepkin, 
Nuovo Cim.\ A \textbf{67}, 211 (1982). 
\bibitem{glueballreferences5}
M.~Strohmeier-Presicek, T.~Gutsche, R.~Vinh Mau and A.~Faessler, 
Phys.\ Rev.\ D \textbf{60}, 054010 (1999) [arXiv:hep-ph/9904461].
\bibitem{glueballreferences6}
C.~Amsler and N.~A.~Tornqvist, 
Phys.\ Rept.\ \textbf{389}, 61 (2004). 
\bibitem{glueballreferences7}
E.~Klempt and A.~Zaitsev, 
Phys.\ Rept.\ \textbf{454}, 1 (2007) [arXiv:0708.4016 [hep-ph]]. 
\bibitem{glueballreferences8} 
V.~Mathieu, N.~Kochelev and V.~Vento,  
Int.\ J.\ Mod.\ Phys.\ E \textbf{18}, 1 (2009)  [arXiv:0810.4453 [hep-ph]].  

\bibitem{Morningstar}  C.~J.~Morningstar and M.~J.~Peardon,  
Phys.\ Rev.\ D \textbf{60}, 034509 (1999)  [hep-lat/9901004].  
\bibitem{Lee}
W.~J.~Lee and D.~Weingarten,  
Phys.\ Rev.\ D \textbf{61}, 014015 (2000)  [hep-lat/9910008].  
 \bibitem{Bali:2000vr} 
G.~S.~Bali \textit{et al.} [TXL and T(X)L Collaborations],  
Phys.\ Rev.\ D \textbf{62}, 054503 (2000)  [hep-lat/0003012].  
\bibitem{Hart} 
A.~Hart \textit{et al.} [UKQCD Collaboration],  
Phys.\ Rev.\ D \textbf{65}, 034502 (2002)  [hep-lat/0108022]. 
\bibitem{Loan}
M.~Loan, X.~Q.~Luo and Z.~H.~Luo, 
Int.\ J.\ Mod.\ Phys.\ A \textbf{21}, 2905 (2006) [arXiv:hep-lat/0503038]. 
\bibitem{Gregory2006}
E.~B.~Gregory, A.~C.~Irving, C.~C.~McNeile, S.~Miller and Z.~Sroczynski, 
PoS \textbf{LAT2005}, 027 (2006) [arXiv:hep-lat/0510066]. 
\bibitem{Chen}
Y.~Chen \textit{et al.}, 
Phys.\ Rev.\ D \textbf{73}, 014516 (2006) [arXiv:hep-lat/0510074]. 
\bibitem{Hart:2006ps} 
A.~Hart \textit{et al.} [UKQCD Collaboration],  
Phys.\ Rev.\ D \textbf{74}, 114504 (2006)  [hep-lat/0608026].  
\bibitem{Richards}
C.~M.~Richards \textit{et al.} [UKQCD Collaboration],  
Phys.\ Rev.\ D \textbf{82}, 034501 (2010)  [arXiv:1005.2473 [hep-lat]].
\bibitem{Gregory}
E.~Gregory, A.~Irving, B.~Lucini, C.~McNeile, A.~Rago, C.~Richards and
E.~Rinaldi,  
JHEP \textbf{1210}, 170 (2012)  [arXiv:1208.1858 [hep-lat]].

\bibitem{Yang}
Y.~B.~Yang {\it et al.} [CLQCD Collaboration],
 Phys.\ Rev.\ Lett.\  {\bf 111}, no. 9, 091601 (2013)
 [arXiv:1304.3807 [hep-lat]].
 
 
 
\bibitem{BMT} 
  R.~C.~Brower, S.~D.~Mathur and C.~I.~Tan,
  Nucl.\ Phys.\ B {\bf 587}, 249 (2000)
  [hep-th/0003115].

\bibitem{Capossoli:2013kb} 
  E.~Folco Capossoli and H.~Boschi-Filho,
  Phys.\ Rev.\ D {\bf 88}, no. 2, 026010 (2013)
  [arXiv:1301.4457 [hep-th]].


\bibitem{Capossoli:2015ywa} 
  E.~Folco Capossoli and H.~Boschi-Filho,
  Phys.\ Lett.\ B {\bf 753}, 419 (2016)
  [arXiv:1510.03372 [hep-ph]].


\bibitem{Capossoli:2016kcr} 
  E.~Folco Capossoli, D.~Li and H.~Boschi-Filho,
  Phys.\ Lett.\ B {\bf 760}, 101 (2016)
  [arXiv:1601.05114 [hep-ph]].


\bibitem{Capossoli:2016ydo} 
  E.~Folco Capossoli, D.~Li and H.~Boschi-Filho,
  Eur.\ Phys.\ J.\ C {\bf 76}, no. 6, 320 (2016)
  [arXiv:1604.01647 [hep-ph]].

 
\bibitem{Hashimoto:2007ze} 
  K.~Hashimoto, C.~I.~Tan and S.~Terashima,
  Phys.\ Rev.\ D {\bf 77}, 086001 (2008)
  [arXiv:0709.2208 [hep-th]].
  
  
\bibitem{Brunner:2015oqa} 
  F.~Br\"{u}nner, D.~Parganlija and A.~Rebhan,
  Phys.\ Rev.\ D {\bf 91}, no. 10, 106002 (2015)
  Erratum: [Phys.\ Rev.\ D {\bf 93}, no. 10, 109903 (2016)]
  [arXiv:1501.07906 [hep-ph]].

  
\bibitem{Brunner:2015yha} 
  F.~Br\"{u}nner and A.~Rebhan,
  Phys.\ Rev.\ Lett.\  {\bf 115}, no. 13, 131601 (2015)
  [arXiv:1504.05815 [hep-ph]].

\bibitem{Brunner:2015oga} 
  F.~Br\"{u}nner and A.~Rebhan,
  Phys.\ Rev.\ D {\bf 92}, no. 12, 121902 (2015)
  [arXiv:1510.07605 [hep-ph]].

  
\bibitem{glueballmodels} F.~Giacosa, T.~Gutsche, V.~E.~Lyubovitskij and
A.~Faessler, 
Phys.\ Rev.\ D \textbf{72}, 094006 (2005) [arXiv:hep-ph/0509247]. 
\bibitem{glueballmodels1}
F.~Giacosa, T.~Gutsche, V.~E.~Lyubovitskij and A.~Faessler, 
Phys.\ Lett.\ B \textbf{622}, 277 (2005) [arXiv:hep-ph/0504033]. 
\bibitem{glueballmodels2}
C.~Amsler and F.~E.~Close, 
Phys.\ Rev.\ D \textbf{53}, 295 (1996) [arXiv:hep-ph/9507326]. 
\bibitem{glueballmodels3}
F.~E.~Close and A.~Kirk, 
Eur.\ Phys.\ J.\ C \textbf{21}, 531 (2001) [arXiv:hep-ph/0103173].
\bibitem{glueballmodels4}
F.~Giacosa, T.~Gutsche and A.~Faessler, 
Phys. Rev. C \textbf{71}, 025202 (2005) [arXiv:hep-ph/0408085].
\bibitem{glueballmodels5}
H.~Y.~Cheng, C.~K.~Chua and K.~F.~Liu, 
Phys.\ Rev.\ D \textbf{74}, 094005 (2006) [arXiv:hep-ph/0607206]. 
\bibitem{Janowski:2011gt} 
S.~Janowski, D.~Parganlija, F.~Giacosa and D.~H.~Rischke,  
Phys.\ Rev.\ D \textbf{84}, 054007 (2011)  [arXiv:1103.3238 [hep-ph]]. 
\bibitem{Janowski:2014ppa} 
S.~Janowski, F.~Giacosa and D.~H.~Rischke,  
Phys.\ Rev.\ D \textbf{90}, no. 11, 114005 (2014)  [arXiv:1408.4921
[hep-ph]].  


\bibitem{Narison:1988ts} 
  S.~Narison and G.~Veneziano,
  Int.\ J.\ Mod.\ Phys.\ A {\bf 4}, 2751 (1989).

\bibitem{Narison:1996fm} 
  S.~Narison,
  Nucl.\ Phys.\ B {\bf 509}, 312 (1998)
  [hep-ph/9612457].


\bibitem{Sanchis-Alepuz:2015hma} 
  H.~Sanchis-Alepuz, C.~S.~Fischer, C.~Kellermann and L.~von Smekal,
  Phys.\ Rev.\ D {\bf 92}, no. 3, 034001 (2015)
  [arXiv:1503.06051 [hep-ph]].

\bibitem{GellMann:1968rz} 
M.~Gell-Mann, R.~J.~Oakes and B.~Renner,
  Phys.\ Rev.\  {\bf 175}, 2195 (1968).

\bibitem{Shifman:1978bw} 
M.~A.~Shifman, A.~I.~Vainshtein and V.~I.~Zakharov,
  Nucl.\ Phys.\ B {\bf 147}, 519 (1979).

\bibitem{Shifman:1978by} 
M.~A.~Shifman, A.~I.~Vainshtein and V.~I.~Zakharov,
  Nucl.\ Phys.\ B {\bf 147}, 448 (1979).

\bibitem{Procura:2006bj} 
  M.~Procura, B.~U.~Musch, T.~Wollenweber, T.~R.~Hemmert and W.~Weise,
  Phys.\ Rev.\ D {\bf 73}, 114510 (2006)
  [hep-lat/0603001].

  
\bibitem{Wiedner:2013bta} 
  U.~Wiedner,
  Acta Phys.\ Polon.\ Supp.\  {\bf 6}, no. 3, 777 (2013).
  
 
\bibitem{Maldacena} 
  J.~M.~Maldacena,
Int.\ J.\ Theor.\ Phys.\  {\bf 38}, 1113 (1999)  [Adv.\ Theor.\ Math.\ Phys.\  {\bf 2}, 231 (1998)]  [hep-th/9711200].  


\bibitem{Witten} 
  E.~Witten,
Adv.\ Theor.\ Math.\ Phys.\  {\bf 2}, 505 (1998)  [hep-th/9803131].  



\bibitem{SaSu1} 
T.~Sakai and S.~Sugimoto,
Prog.\ Theor.\ Phys.\ \textbf{113}, 843 (2005) [hep-th/0412141].

\bibitem{SaSu2} 
T.~Sakai and S.~Sugimoto,
Prog.\ Theor.\ Phys.\ \textbf{114}, 1083 (2005) [hep-th/0507073].


\bibitem{West} 
  G.~B.~West,
  Phys.\ Rev.\ Lett.\  {\bf 77}, 2622 (1996)
  [hep-ph/9603316].
  

\bibitem{PDG} K.~A.~Olive {\it et al.} [Particle Data Group Collaboration],
  Chin.\ Phys.\ C {\bf 38}, 090001 (2014) and 2015 update.

\bibitem{BES} 
  J.~F.~Liu {\it et al.} [BES Collaboration],
  Phys.\ Rev.\ D {\bf 82}, 074026 (2010)
  [arXiv:1008.0246 [hep-ph]].
  
\bibitem{Belle} 
  M.-Z.~Wang {\it et al.} [Belle Collaboration],
  Phys.\ Lett.\ B {\bf 617}, 141 (2005)
  [hep-ex/0503047].

\bibitem{LHCb} 
  A.~Palano [LHCb Collaboration],
  Acta Phys.\ Polon.\ Supp.\  {\bf 8}, no. 1, 159 (2015).


  
\bibitem{Bugg:2004xu}  D.~V.~Bugg,  
Phys.\ Rept.\ \textbf{397}, 257 (2004)  [hep-ex/0412045].  


\bibitem{Parganlija:2012nc} 
  D.~Parganlija,
  J.\ Phys.\ Conf.\ Ser.\  {\bf 426}, 012019 (2013)
  [arXiv:1211.4804 [hep-ph]].
  
  
\bibitem{Parganlija:2013xsa} 
  D.~Parganlija,
  J.\ Phys.\ Conf.\ Ser.\  {\bf 503}, 012010 (2014)
  [arXiv:1312.2830 [hep-ph]].
  
  
  
\bibitem{1790-1} A.~V.~Anisovich \textit{et al.}, 
Phys.\ Lett.\ B \textbf{449}, 154 (1999). 
\bibitem{1790-2}
B.~S.~Zou, 
Nucl.\ Phys.\ A \textbf{692}, 362 (2001) [arXiv:hep-ph/0011174]. 


\bibitem{BESII2004-1} M.~Ablikim \textit{et al.} [BES Collaboration], 
Phys.\ Lett.\ B \textbf{603}, 138 (2004) [arXiv:hep-ex/0409007].

\bibitem{BESII2004-2}
M.~Ablikim \textit{et al.} [BES Collaboration], 
Phys.\ Lett.\ B \textbf{607}, 243 (2005) [arXiv:hep-ex/0411001]. 


\bibitem{DA} 
  D.~Parganlija,
  arXiv:1208.0204 [hep-ph].
  
  
\bibitem{1710} 
D.~Barberis \textit{et al.} [WA102 Collaboration], 
Phys.\ Lett.\ B \textbf{462}, 462 (1999) [arXiv:hep-ex/9907055]. 


\bibitem{NICA} 
  V.~Kekelidze {\it et al.} [NICA Collaboration],
  EPJ Web Conf.\  {\bf 95}, 01014 (2015).
  
  
\bibitem{SPD} 
  I.~Savin {\it et al.},
  EPJ Web Conf.\  {\bf 85}, 02039 (2015).

\bibitem{SPD1}
I.~Savin {\it et al.},
Eur.\ Phys.\ J.\ A {\bf 52}, 215 (2016);
 see also
  "Spin Physics Experiments at NICA-SPD with polarized proton and deuteron beam. Letter of Intent."
  (NICA LoI-02.06.14).
  
\bibitem{BALDIN:2014wia} 
  E.~M.~Baldin [KEDR Collaboration],
  PoS EPS {\bf -HEP2013}, 326 (2013).

\bibitem{Schwarz:2014nfa} 
  B.~A.~Shwartz [CMD-3 and SND Collaborations],
  PoS Hadron {\bf 2013}, 019 (2013).

\bibitem{Abe:2010gxa} 
  T.~Abe {\it et al.} [Belle-II Collaboration],
  arXiv:1011.0352 [physics.ins-det].
  
\bibitem{Anelli:2008zza} 
  G.~Anelli {\it et al.} [TOTEM Collaboration],
  JINST {\bf 3}, S08007 (2008).

  
\bibitem{Lamsa:2010fe} 
  J.~W.~Lamsa and R.~Orava,
  JINST {\bf 6}, P02010 (2011)
  [arXiv:1009.3350 [hep-ex]].

\bibitem{PANDA} 
  M.~F.~M.~Lutz {\it et al.} [PANDA Collaboration],
  arXiv:0903.3905 [hep-ex].
  
\bibitem{COMPASS}
  B.~Grube [COMPASS Collaboration],
  AIP Conf.\ Proc.\  {\bf 1735}, 020007 (2016)
  [arXiv:1512.03599 [hep-ex]].
  
\bibitem{GlueX} 
  C.~A.~Meyer {\it et al.} [GlueX Collaboration],
  AIP Conf.\ Proc.\  {\bf 1735}, 020001 (2016)
  [arXiv:1512.03699 [nucl-ex]].
  
\bibitem{CLAS12} 
  D.~I.~Glazier,
  Acta Phys.\ Polon.\ Supp.\  {\bf 8}, no. 2, 503 (2015).

  
\bibitem{Divotgey:2013jba} 
  F.~Divotgey, L.~Olbrich and F.~Giacosa,
  Eur.\ Phys.\ J.\ A {\bf 49}, 135 (2013)
  [arXiv:1306.1193 [hep-ph]].
  
\bibitem{Parganlija:2010fz} 
D.~Parganlija, F.~Giacosa and D.~H.~Rischke,
Phys.\ Rev.\ D {\bf 82}, 054024 (2010)  [arXiv:1003.4934 [hep-ph]].
  
  
\bibitem{Parganlija:2012fy} 
D.~Parganlija, P.~Kovacs, G.~Wolf, F.~Giacosa and D.~H.~Rischke,
Phys.\ Rev.\ D {\bf 87}, 014011 (2013)  [arXiv:1208.0585 [hep-ph]].


\bibitem {RS}S.~Str\"{u}ber and D.~H.~Rischke,
Phys.\ Rev.\ D \textbf{77}, 085004 (2008) [arXiv:0708.2389 [hep-th]].

\bibitem{BR} G.~E.~Brown and M.~Rho, 
Phys.\ Rev.\ Lett.\ \textbf{66}, 2720 (1991). 

\bibitem{Pelaez:2006nj} 
  J.~R.~Pelaez and G.~Rios,
  Phys.\ Rev.\ Lett.\  {\bf 97}, 242002 (2006)
  [hep-ph/0610397].



  
\bibitem{Boyd:1996bx} 
  G.~Boyd, J.~Engels, F.~Karsch, E.~Laermann, C.~Legeland, M.~Lutgemeier and B.~Petersson,
  Nucl.\ Phys.\ B {\bf 469}, 419 (1996)
  [hep-lat/9602007].
  
\bibitem{Boyd:1996ex} 
  G.~Boyd and D.~E.~Miller,
  hep-ph/9608482.


\bibitem{Miller:2000ki} 
  D.~E.~Miller,
  hep-ph/0008031.

\bibitem{Miller:2006hr} 
  D.~E.~Miller,
  Phys.\ Rept.\  {\bf 443}, 55 (2007)
  [hep-ph/0608234].

\bibitem{Sollfrank:1994du} 
  J.~Sollfrank and U.~W.~Heinz,
  Z.\ Phys.\ C {\bf 65}, 111 (1995)
  [nucl-th/9406014].

  
\bibitem{Agasian:1997zr} 
  N.~O.~Agasian, D.~Ebert and E.~M.~Ilgenfritz,
  Nucl.\ Phys.\ A {\bf 637}, 135 (1998)
  [hep-ph/9712344].

  
\bibitem{Sasaki:2011sd} 
  C.~Sasaki and I.~Mishustin,
  Phys.\ Rev.\ C {\bf 85}, 025202 (2012)
  [arXiv:1110.3498 [hep-ph]].

  
\bibitem{Agasian:1993fn} 
  N.~O.~Agasian,
  JETP Lett.\  {\bf 57}, 208 (1993)
  [Pisma Zh.\ Eksp.\ Teor.\ Fiz.\  {\bf 57}, 200 (1993)].

  
\bibitem{Volkov:1994eu} 
  M.~K.~Volkov,
  Theor.\ Math.\ Phys.\  {\bf 101}, 1473 (1994)
  [Teor.\ Mat.\ Fiz.\  {\bf 101}, 442 (1994)].

  
\bibitem{Schaefer:2001cn} 
  B.~J.~Schaefer, O.~Bohr and J.~Wambach,
  Phys.\ Rev.\ D {\bf 65}, 105008 (2002)
  [hep-th/0112087].


\bibitem{ELSA} 
  D.~Trnka {\it et al.} [CBELSA/TAPS Collaboration],
Phys.\ Rev.\ Lett.\  {\bf 94}, 192303 (2005)  [nucl-ex/0504010].  

\bibitem{KEK1} 
  M.~Naruki {\it et al.},
Phys.\ Rev.\ Lett.\  {\bf 96}, 092301 (2006)  [nucl-ex/0504016].  



\bibitem{KEK2} 
  M.~Naruki {\it et al.} [E325 Collaboration],
J.\ Phys.\ G {\bf 34}, S1059 (2007).  


\bibitem{CLAS} 
  M.~H.~Wood {\it et al.} [CLAS Collaboration],
Phys.\ Rev.\ C {\bf 78}, 015201 (2008)  [arXiv:0803.0492 [nucl-ex]].  

\bibitem{PHENIX} 
  Y.~Tsuchimoto [PHENIX Collaboration],
Nucl.\ Phys.\ A {\bf 830}, 487C (2009)  [arXiv:0907.5049 [hep-ex]].  

\bibitem{ANKE} 
  A.~Polyanskiy {\it et al.},
Phys.\ Lett.\ B {\bf 695}, 74 (2011)  [arXiv:1008.0232 [nucl-ex]].  



\bibitem{MAMI-C} 
  M.~Thiel {\it et al.},
Eur.\ Phys.\ J.\ A {\bf 49}, 132 (2013).  


\bibitem{McLerran:2009ry} 
  L.~McLerran,
  arXiv:0911.2987 [hep-ph].

  
\bibitem{McLerran:2014hza} 
  L.~McLerran and B.~Schenke,
  Nucl.\ Phys.\ A {\bf 929}, 71 (2014)
  [arXiv:1403.7462 [hep-ph]].

  
\bibitem{Akiba:2015jwa} 
  Y.~Akiba {\it et al.},
  arXiv:1502.02730 [nucl-ex].

  
\bibitem{Buisseret:2009eb} 
  F.~Buisseret,
  Eur.\ Phys.\ J.\ C {\bf 68}, 473 (2010)
  [arXiv:0912.0678 [hep-ph]].

  
\bibitem{Ishii:2002ww}
  N.~Ishii, H.~Suganuma and H.~Matsufuru,
  Phys.\ Rev.\ D {\bf 66}, 094506 (2002)
  [hep-lat/0206020].

  
\bibitem{Meng:2009hh} 
  X.~F.~Meng, G.~Li, Y.~Chen, C.~Liu, Y.~B.~Liu, J.~P.~Ma and J.~B.~Zhang,
  Phys.\ Rev.\ D {\bf 80}, 114502 (2009)
  [arXiv:0903.1991 [hep-lat]].

  

\bibitem{Lacroix:2012pt} 
  G.~Lacroix, C.~Semay, D.~Cabrera and F.~Buisseret,
  Phys.\ Rev.\ D {\bf 87}, no. 5, 054025 (2013)
 [arXiv:1210.1716 [hep-ph]].

\bibitem{Kochelev:2015rqa} 
  N.~Kochelev,
  Phys.\ Part.\ Nucl.\ Lett.\  {\bf 13}, no. 2, 149 (2016)
  [arXiv:1501.07002 [hep-ph]].


\bibitem{Vento:2006wh} 
  V.~Vento,
  Phys.\ Rev.\ D {\bf 75}, 055012 (2007)
  [hep-ph/0609219].

  
\bibitem{Kochelev:2006sx} 
  N.~Kochelev and D.~P.~Min,
  Phys.\ Lett.\ B {\bf 650}, 239 (2007)
  [hep-ph/0611250].

  
  
\bibitem{MPD} 
  V.~D.~Kekelidze {\it et al.} [NICA and MPD Collaborations],
  Phys.\ Atom.\ Nucl.\  {\bf 75}, 542 (2012).

\bibitem{MPD1}
 V.~Golovatyuk, V.~Kekelidze, V.~Kolesnikov, O.~Rogachevsky and A.~Sorin,
  Eur.\ Phys.\ J.\ A {\bf 52}, no. 8, 212 (2016);
see also
"The MultiPurpose Detector -- MPD to study Heavy Ion Collisions at NICA
(Conceptual Design Report)"
(JINR, Dubna, Version 1.4).
 
  
\bibitem{BMN} 
  V.~Ladygin [BM@N Collaboration],
  PoS Baldin {\bf -ISHEPP-XXI}, 038 (2012) and refs.\ therein.

\bibitem{BMN1} 
  M.~Kapishin [BM@N Collaboration],
  Eur.\ Phys.\ J.\ A {\bf 52}, no. 8, 213 (2016).

  
\bibitem{RHIC} 
  G.~S.~F.~Stephans,
  J.\ Phys.\ G {\bf 32}, S447 (2006)
  [nucl-ex/0607030].

  
\bibitem{ALICE} 
  R.~Tieulent [ALICE Collaboration],
  arXiv:1512.02253 [nucl-ex].

  
\bibitem{HADES} 
  P.~Salabura {\it et al.} [HADES Collaboration],
  J.\ Phys.\ Conf.\ Ser.\  {\bf 420}, 012013 (2013).

  
\bibitem{SPS} 
  B.~Lungwitz {\it et al.} [NA49 Collaboration],
  PoS CPOD {\bf 07}, 023 (2007)
  [arXiv:0709.1646 [nucl-ex]].

  
\bibitem{CBM} 
  P.~Senger,
  J.\ Phys.\ G {\bf 30}, S1087 (2004).

  
\bibitem{Hohler:2013eba} 
  P.~M.~Hohler and R.~Rapp,
  Phys.\ Lett.\ B {\bf 731}, 103 (2014)
  [arXiv:1311.2921 [hep-ph]].
  
\bibitem{Hohler:2015iba} 
  P.~M.~Hohler and R.~Rapp,
  Annals Phys.\  {\bf 368}, 70 (2016)
  [arXiv:1510.00454 [hep-ph]].


\bibitem{Kovacs:2016juc} 
  P.~Kov{\'a}cs, Z.~Sz{\'e}p and G.~Wolf,
  Phys.\ Rev.\ D {\bf 93}, no. 11, 114014 (2016)
  [arXiv:1601.05291 [hep-ph]].

  
\bibitem{Glozman:2003bt} 
  L.~Y.~Glozman,
  Phys.\ Lett.\ B {\bf 587}, 69 (2004)
  [hep-ph/0312354].
  

\bibitem{Glozman:2007ek} 
  L.~Y.~Glozman,
  Phys.\ Rept.\  {\bf 444}, 1 (2007)
  [hep-ph/0701081].

  


  
\end{thebibliography}

\end{document}